\documentclass[preprint,12pt]{elsarticle}

\usepackage{color,longtable,hyperref}
\usepackage{xspace}
\usepackage{amsmath}
\usepackage{axodraw}

\definecolor{mygreen}{rgb}{0,1,0}
\definecolor{mygreen}{rgb}{0,.75,0}

\definecolor{mycyan}{cmyk}{1,0,0,0}
\definecolor{mycyan}{cmyk}{.8,.15,0,0}
\definecolor{mycyan}{cmyk}{.8,.55,0,0}

\definecolor{mymagenta}{cmyk}{0,1,0,0}
\definecolor{mymagenta}{cmyk}{.15,1,0,0}

\usepackage{amssymb}

\biboptions{compress}

\newcounter{bla}

\journal{Computer Physics Communications}

\newcommand{\Prophecy}{{\sc Pro\-phecy4f}\xspace}
\newcommand{\Collier}{{\sc Collier}\xspace}

\def\refeq#1{\mbox{(\ref{#1})}}
\def\reffi#1{\mbox{Fig.~\ref{#1}}}

\def\refse#1{\mbox{Section~\ref{#1}}}
\def\refses#1{\mbox{Sections~\ref{#1}}}

\def\citere#1{\mbox{Ref.~\cite{#1}}}
\def\citeres#1{\mbox{Refs.~\cite{#1}}}

\def\mathswitch#1{\relax\ifmmode#1\else$#1$\fi}
\def\mathswitchr#1{\relax\ifmmode{\mathrm{#1}}\else$\mathrm{#1}$\fi}
\def\mathswitchit#1{\relax\ifmmode{#1}\else$#1$\fi}

\newcommand{\PV}{\mathswitch V}
\newcommand{\PW}{\mathswitchr W}
\newcommand{\PZ}{\mathswitchr Z}

\newcommand{\PH}{\mathswitchr H}

\newcommand{\Pe}{\mathswitchr e}

\newcommand{\Pd}{\mathswitchr d}
\newcommand{\Pf}{f}
\newcommand{\Ph}{\mathswitchr h}

\newcommand{\Pu}{\mathswitchr u}
\newcommand{\Ps}{\mathswitchr s}

\newcommand{\Pc}{\mathswitchr c}

\newcommand{\Pep}{\mathswitchr {e^+}}
\newcommand{\Pem}{\mathswitchr {e^-}}

\newcommand{\MW}{\mathswitch {M_\PW}}

\newcommand{\MZ}{\mathswitch {M_\PZ}}

\newcommand{\GeV}{\unskip\,\mathrm{GeV}}

\newcommand{\MSbar}{\ensuremath{\overline{\mathrm{MS}}}\xspace}

\newenvironment{cpcdescription}
   {\begin{description}%
   \setlength{\itemsep}{2ex}}%
   {\end{description}}

\newlength{\colonewidth}
\newlength{\parwidth}

\newcommand{\cpcitemtable}[2]
{\settowidth{\colonewidth}{#1}\setlength{\parwidth}{\textwidth}%
\addtolength{\parwidth}{-\leftmargin}%
\addtolength{\parwidth}{-\colonewidth}\addtolength{\parwidth}{-2em}%
\nobreak
\begin{flushleft}%
\vspace{-1.5ex}
\begin{tabular}[l]{@{}p{\colonewidth}@{ }c@{ }p{\parwidth}}%
#2 
\end{tabular}%
\vspace{-1.ex}
\end{flushleft}%
\ignorespaces}

\newcommand{\cpctable}[2]
{\settowidth{\colonewidth}{#1}\setlength{\parwidth}{\textwidth}%
\addtolength{\parwidth}{-\leftmargin}%
\addtolength{\parwidth}{-\colonewidth}\addtolength{\parwidth}{-2em}%
\nobreak
\begin{longtable}[l]{@{\qquad}p{\colonewidth}@{ }c@{ }p{\parwidth}}%
#2 
\end{longtable}%
\noindent\ignorespaces}

\def\draftdate{\relax}
\def\mda{\relax}
\def\mua{\relax}
\def\mla{\relax}
\def\draft{
\def\thtystars{******************************}
\def\sixtystars{\thtystars\thtystars}
\typeout{}
\typeout{\sixtystars**}
\typeout{* Draft mode!
         For final version remove \protect\draft\space in source file *}
\typeout{\sixtystars**}
\typeout{}
\def\draftdate{\today}
\def\mua{\marginpar[\boldmath\hfil$\uparrow$]%
                   {\boldmath$\uparrow$\hfil}%
                    \typeout{marginpar: $\uparrow$}\ignorespaces}
\def\mda{\marginpar[\boldmath\hfil$\downarrow$]%
                   {\boldmath$\downarrow$\hfil}%
                    \typeout{marginpar: $\downarrow$}\ignorespaces}
\def\mla{\marginpar[\boldmath\hfil$\rightarrow$]%
                   {\boldmath$\leftarrow $\hfil}%
                    \typeout{marginpar: $\leftrightarrow$}\ignorespaces}
\def\Mua{\marginpar[\boldmath\hfil$\Uparrow$]%
                   {\boldmath$\Uparrow$\hfil}%
                    \typeout{marginpar: $\Uparrow$}\ignorespaces}
\def\Mda{\marginpar[\boldmath\hfil$\Downarrow$]%
                   {\boldmath$\Downarrow$\hfil}%
                    \typeout{marginpar: $\Downarrow$}\ignorespaces}
\def\Mla{\marginpar[\boldmath\hfil$\Rightarrow$]%
                   {\boldmath$\Leftarrow $\hfil}%
                    \typeout{marginpar: $\Leftrightarrow$}\ignorespaces}
\overfullrule 5pt
\oddsidemargin -15mm
\marginparwidth 29mm
}

\begin{document}

\begin{frontmatter}



\title{\Prophecy\ 3.0: A Monte Carlo program for Higgs-boson decays
into four-fermion final states\\ in and beyond the Standard Model}


\author[a]{Ansgar Denner}
\author[b]{Stefan Dittmaier}
\author[c]{Alexander M\"uck\corref{author}}

\cortext[author] 
{Corresponding author.\\\textit{E-mail address:} mueck@physik.rwth-aachen.de}
\address[a]{Universit\"at W\"urzburg, Institut f\"ur Theoretische Physik und Astrophysik, \\
D-97074 W\"urzburg, Germany}
\address[b]{Albert-Ludwigs-Universit\"at Freiburg, Physikalisches Institut, \\
D-79104 Freiburg, Germany}
\address[c]{RWTH Aachen University, Institut f\"ur Theoretische Teilchenphysik und Kosmologie, \\
D-52056 Aachen, Germany}

\begin{abstract}
  The Monte Carlo generator \Prophecy\ provides a \underline{PROP}er
  description of the \underline{H}iggs d\underline{EC}a\underline{Y}
  into \underline{4} \underline{F}ermions within the Standard Model,
  the Standard Model with a fourth fermion generation, a simple
  Higgs-singlet extension of the Standard Model, and the
  Two-Higgs-Doublet Model. The fully differential predictions include
  the full QCD and electroweak next-to-leading-order corrections, all
  interference contributions between different WW/ZZ channels, and all
  off-shell effects of intermediate W/Z~bosons. \Prophecy\ computes
  the inclusive partial decay widths and allows for the computation of
  binned differential distributions of the decay products. For
  leptonic final states also unweighted events are provided. 
\end{abstract}

\begin{keyword}
Higgs physics, radiative corrections, Monte Carlo integration

\end{keyword}

\end{frontmatter}



\noindent
{\bf PROGRAM SUMMARY}\\

\begin{small}
\noindent
{\em Manuscript Title:}                                       
\Prophecy\ 3.0: A Monte Carlo program for Higgs-boson decays
into four-fermion final states in and beyond the Standard Model  \\[1ex]
{\em Authors:}
Ansgar Denner, 
Stefan Dittmaier,
Alexander M\"uck\\[1ex]
{\em Program Title:} 
\Prophecy, version 3.0                                          \\[1ex]
{\em Journal Reference:}                                      \\[1ex]
{\em Catalogue identifier:}                                   \\[1ex]
{\em Licensing provisions:}
none                                   \\[1ex]
{\em Programming language:}
Fortran 77, Fortran 95                                   \\[1ex]
{\em Computer:}
Any computer with a Fortran 95 compiler                          \\[1ex]
{\em Operating system:}
Linux, Mac OS                                       \\[1ex]
{\em RAM:} less than 1 GB                                              \\[1ex]
{\em Keywords:}  
Higgs physics, radiative corrections, Monte Carlo integration \\[1ex]
{\em Classification:}
4.4 Feynman Diagrams,
11.1 General, High Energy Physics and Computing,
11.2 Phase Space and Event Simulation.                 \\[1ex]
{\em External routines/libraries:} 
\Collier (\href{https://collier.hepforge.org/}{\url{https://collier.hepforge.org/}}).      \\[1ex]
{\em Nature of problem:} 
Precision calculation of partial decay widths and differential
distributions for Higgs-boson decays into four-fermion final states as described
in~\citeres{Bredenstein:2006rh,Bredenstein:2006nk,Bredenstein:2006ha,
  Altenkamp:2017ldc,Altenkamp:2017kxk,Altenkamp:2018bcs,Denner:2018opp}.\\[1ex]
{\em Solution method:}
Multi-channel Monte Carlo integration of perturbative matrix elements
including higher-order QCD and electroweak corrections which are based
on a Feynman-diagrammatic calculation.
\\[1ex]
{\em Restrictions:}
No unweighted events for semileptonic and hadronic final states.
\\[1ex]
{\em Running time:}
For $10^7$ weighted events the program will roughly run $10$--$90$ minutes
depending on the final state, hardware, and compilers. The production
of $10^6$ unweighted events will take about $1$--$2$ days.

%
\end{small}%


\section{Introduction}

Presently, one of the main objectives of the experiments at CERN's
Large Hadron Collider (LHC) is the precise investigation of the Higgs
boson. Pinning down its nature and searching for 
deviations from
the predictions of the Standard Model (SM) requires precise
theoretical predictions.  In a coordinated effort between the theory
community and the LHC experiments, theoretical predictions and
recommendations are compiled by the LHC Higgs Cross Section Working
Group (LHCHXSWG)
~\cite{Dittmaier:2011ti,Dittmaier:2012vm,Heinemeyer:2013tqa,deFlorian:2016spz}.

A crucial role for the accurate investigation of the properties of the
Higgs boson is played by the Higgs branching ratios and decay widths.
The SM Higgs-boson branching ratios used by the LHC experiments are
provided by the
LHCHXSWG~\cite{Dittmaier:2011ti,Dittmaier:2012vm,Heinemeyer:2013tqa,deFlorian:2016spz}
based on the codes {\sc HDECAY}
\cite{Djouadi:1997yw,Djouadi:2006bz,Djouadi:2018xqq} for the two-particle
decays and \Prophecy
\cite{Bredenstein:2006rh,Bredenstein:2006nk,Bredenstein:2006ha} for
the Higgs decays into four fermions. While both codes were initially
constructed for the SM, they have been extended to models beyond.
{Recent versions~\cite{Djouadi:2018xqq,Krause:2018wmo} 
of {\sc HDECAY} include the SM with four generations of fermions, a general
Two-Higgs-Doublet Model (THDM),}
and the minimal supersymmetric
Standard Model (MSSM). Moreover, an extension to the SM effective
field theory exists~\cite{Contino:2014aaa}.

\Prophecy is a state-of-the-art tool to predict the SM Higgs-boson
decay widths into four arbitrary fermions (i.e.\ quarks or leptons)
via W and/or Z pairs, including full electroweak (EW) and QCD
next-to-leading order (NLO) corrections.  These processes {have}
played a central role in the discovery of the Higgs boson, and are
very important channels in Higgs coupling analyses {(see
  \citeres{Heinemeyer:2013tqa,deFlorian:2016spz} and references
  therein).  Originally, \Prophecy had been designed for Higgs-boson
  decays in the SM, but in recent years it has been extended to the
  Standard Model with a fourth fermion generation 
  and to the corresponding decays of CP-even, neutral Higgs bosons in
  the THDM \cite{Gunion:2002zf,Branco:2011iw} and in a simple
  Higgs-singlet extension of the SM (SESM) \cite{Schabinger:2005ei,Patt:2006fw,Bowen:2007ia},
  as described in \citeres{Denner:2011vt},
  \cite{Altenkamp:2017ldc,Altenkamp:2017kxk}, and
  \cite{Altenkamp:2018bcs}, respectively.
As a particular strength of our THDM and SESM implementations,
\Prophecy supports many different renormalization schemes, even
different types of schemes based on $\MSbar$, on-shell, or
symmetry-inspired renormalization conditions~\cite{Denner:2018opp}.
This allows for an assessment of uncertainties in predictions due to
residual renormalization scale and renormalization scheme dependences.
We consider this feature important, since many examples in the
literature have shown that individual renormalization schemes might
produce unreliable results in specific regions of parameter space of
model extensions (see, e.g.,
\citeres{Altenkamp:2017ldc,Altenkamp:2017kxk,Denner:2018opp} for such
examples of $\PH\to4f$ decays in the THDM).  On the technical side, the
internal library of \Prophecy for one-loop integrals is now replaced
by the public integral library \Collier
\cite{collier,Denner:2016kdg}.}

An alternative to \Prophecy for the simulation of the SM Higgs-boson
decays into four charged leptons is {the Monte Carlo} event
generator {\sc Hto4L} \cite{Boselli:2015aha}. Besides
the complete NLO EW corrections, it includes also multiple-photon effects in a
matched-to-NLO Parton Shower framework.  The SM results of {\sc Hto4L}
are fully equivalent to those of \Prophecy for the inclusive partial
decay widths and branching ratios.  The effects of multi-photon
emissions on distributions provided by {\sc Hto4L} are typically below
one percent.  The code {\sc Hto4L} has been extended to the SM
effective field theory \cite{Boselli:2017pef}.

The purpose of this paper is to describe version 3.0 of
\Prophecy including all of its extensions and to document its use.

This paper is organized as follows: In \refse{se:details} we describe
the basic features of \Prophecy and highlight some important details.
Section \ref{se:usage} provides all information how to use and run the
code, and \refse{se:output} describes the output of the code and the
sample runs. Concluding remarks are contained in \refse
{se:conclusion}.

\section{\Prophecy description}
\label{se:details}

\Prophecy provides predictions for decays of on-shell Higgs bosons via
a pair of virtual W/Z bosons into four fermions in the SM, the SM with
a fourth fermion generation, a simple Higgs-singlet extension of the
SM, and the Two-Higgs-Doublet Model.  External fermions are considered
in the massless limit.  In the non-standard models, \Prophecy only
deals with the decays of the CP-even, neutral Higgs bosons, and
possible four-fermion decay channels via Higgs-boson pairs are not
considered (in line with the massless limit of the external fermions).
\Prophecy produces fully
differential predictions in the Higgs-boson rest frame, including the
full QCD and electroweak next-to-leading-order corrections, all
interference contributions between different WW/ZZ channels at leading
order (LO) and NLO, and all off-shell effects of intermediate
W/Z~bosons. For decays within the SM also an improved Born
approximation (IBA) is implemented.

\subsection{Process definition}

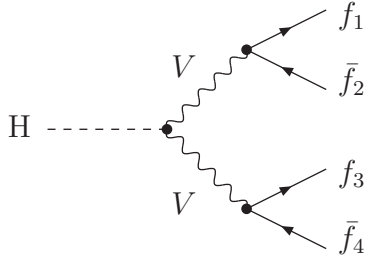
\begin{figure}
\begin{center}
\setlength{\unitlength}{1pt}
\begin{picture}(140,100)(0,0)
\DashLine(15,50)(60,50){3}
\Photon(60,50)(90,20){-2}{5}
\Photon(60,50)(90,80){2}{5}
\Vertex(60,50){2.0}
\Vertex(90,80){2.0}
\Vertex(90,20){2.0}
\ArrowLine(90,80)(120, 95)
\ArrowLine(120,65)(90,80)
\ArrowLine(120, 5)( 90,20)
\ArrowLine( 90,20)(120,35)
\put(0,47){$\PH$}
\put(62,70){$V$}
\put(62,18){$V$}
\put(125,90){$f_1$}
\put(125,65){$\bar f_2$}
\put(125,30){$f_3$}
\put(125,5){$\bar f_4$}
\end{picture}
\end{center}
\caption{Generic lowest-order diagram for $\PH\to 4f$ where $V=\PW,\PZ$.}
\label{fi:H4f-born-diag}
\end{figure}

The considered processes are of the form
\begin{equation}\label{eq:process}
\PH(p) \to f_1(k_1) +  \bar{f}_2(k_2) +  f_3(k_3) +  \bar{f}_4(k_4),
\end{equation}
where the momenta are indicated in parentheses.  The generic LO
Feynman diagram is shown in \reffi{fi:H4f-born-diag}.  The external
fermions are assumed to be massless; their masses only serve as
regulators in case of logarithmic mass singularities, which arise for
instance if final-state fermions and photons are not recombined.  
While the order of fermions in the input is arbitrary, 
{the produced distributions are based on a
canonical order of the final-state fermions in \refeq{eq:process}, where the
fermion--antifermion pairs $f_1\bar{f_2}$ and $f_3\bar{f_4}$ are
related to the virtual W/Z bosons. For processes, where both virtual
WW and ZZ pairs are possible, $f_1\bar{f_2}$ and $f_3\bar{f_4}$ are
associated to the W bosons, as further explained in \refse{se:Differential}.}

\subsection{EW input-parameter scheme}

For the electromagnetic coupling constant, we use the $G_\mathrm{F}$
scheme, i.e.\ the coupling constant $\alpha$ is derived from the Fermi
constant according to
\begin{equation}
 \alpha_{G_\mathrm{F}} = \frac{\sqrt{2}G_\mathrm{F}\MW^2}{\pi}\left(1-\frac{\MW^2}{\MZ^2}\right) .
\end{equation}
This procedure takes into account some higher-order effects, related
to the running of the electromagnetic coupling and to the $\rho$
parameter, already at tree level.

\subsection{Complex-mass scheme,  input masses and widths}
\label{sec:cms}

Gauge-boson resonances are treated using the complex-mass scheme
\cite{Denner:1999gp,Denner:2005fg}, i.e.\ all corresponding decay and
off-shell effects are supported at NLO accuracy in the full phase
space.  As input, the program expects the on-shell W/Z masses and
widths.  From these it calculates internally the real parts of the
complex pole masses using
\begin{equation}
\label{eq:m_ga_pole}
M_V^{\mathrm{pole}} = \frac{M_{\PV}^{\mbox{\scriptsize{on-shell}}}}
{\sqrt{1+(\Gamma_{\PV}^{\mbox{\scriptsize{on-shell}}}/M_{\PV}^{\mbox{\scriptsize{on-shell}}})^2}}\;.
\end{equation}
The imaginary parts of the complex pole masses, i.e.\ the vector-boson
widths are calculated as described below. These pole masses are then
used in propagators{, the} complex weak mixing angle{, and} other
couplings for the evaluation of the Higgs-boson decays.

The decay widths of the W and Z bosons that enter the complex pole
masses are calculated from the pole masses and the other input
parameters as follows: If only LO results are requested, the LO
widths are used. For NLO results and the IBA we employ the NLO widths
(also for the LO sub-contribution). This ensures that the effective
branching fractions for the W- and Z-boson decays in both LO and NLO
add up to one. 

Since the NLO W and Z widths in the SESM and the THDM are almost
identical to the corresponding SM NLO widths, the SM widths are used
employing a SM Higgs mass set to the mass of the decaying Higgs boson.

\subsection{Total width}
The total width is calculated according to 
\begin{align}\label{eq:total_width}
\Gamma_{\PH\to4f} ={}& \Gamma^{\mathrm{total}}
= \Gamma^{\mathrm{leptonic}} + \Gamma^{\mbox{\scriptsize semi-leptonic}} +\Gamma^{\mathrm{hadronic}}
\end{align}
with
\begin{align}
\Gamma^{\mathrm{leptonic}} ={}& 
  3\Gamma^{\nu_\Pe \bar{\nu}_\Pe \nu_\mu \bar{\nu}_\mu}
+ 3\Gamma^{\Pem\Pep\mu^-\mu^+}
+ 6\Gamma^{\nu_\Pe \bar{\nu}_\Pe \mu^-\mu^+}
+ 6\Gamma^{\nu_\Pe \Pep \mu^-\bar{\nu}_\mu}
\notag\\&
+ 3\Gamma^{\nu_\Pe \bar{\nu}_\Pe \nu_\Pe \bar{\nu}_\Pe}
+ 3\Gamma^{\Pem\Pep\Pem\Pep}
+ 3\Gamma^{\nu_\Pe \Pep\Pem \bar{\nu}_\Pe},
\notag\\[1ex]
\Gamma^{\mathrm{hadronic}} ={}& 
   \Gamma^{\Pu\bar{\Pu}\Pc\bar{\Pc}}
+ 3\Gamma^{\Pd\bar{\Pd}\Ps\bar{\Ps}}
+ 4\Gamma^{\Pu\bar{\Pu}\Ps\bar{\Ps}}
+ 2\Gamma^{\Pu\bar{\Pd}\Ps\bar{\Pc}}
+ 2\Gamma^{\Pu\bar{\Pu}\Pu\bar{\Pu}}
+ 3\Gamma^{\Pd\bar{\Pd}\Pd\bar{\Pd}}
+ 2\Gamma^{\Pu\bar{\Pd}\Pd\bar{\Pu}},
\notag\\[1ex]
\Gamma^{\mbox{\scriptsize semi-leptonic}} ={}& 
  6\Gamma^{\nu_\Pe \bar{\nu}_\Pe \Pu\bar{\Pu}}
+ 9\Gamma^{\nu_\Pe \bar{\nu}_\Pe \Pd\bar{\Pd}}
+ 6\Gamma^{\Pu\bar{\Pu}\Pem\Pep}
+ 9\Gamma^{\Pd\bar{\Pd}\Pem\Pep}
+ 12\Gamma^{\nu_\Pe \Pep \Pd\bar{\Pu}}.
\label{eq:subtotal_width}
\end{align}

The total width can be split into decays via ZZ, WW and the
interference,
\begin{equation}
\Gamma_{\PH\to4f} =
\Gamma_{\PH\to\PW^*\PW^*\to4f}
+\Gamma_{\PH\to\PZ^*\PZ^*\to4f}
+\Gamma_{\PW\PW/\PZ\PZ\mbox{\scriptsize -interference}},
\end{equation}
where the individual terms are defined in terms of partial widths with
specific final states as
\begin{align}
\Gamma_{\PH\to\PW^*\PW^*\to4f} ={}& 
9\Gamma^{\nu_\Pe \Pep \mu^-\bar{\nu}_\mu}
+ 12\Gamma^{\nu_\Pe \Pep \Pd\bar{\Pu}}
+ 4\Gamma^{\Pu\bar{\Pd}\Ps\bar{\Pc}},
\notag\\[1ex]
\Gamma_{\PH\to\PZ^*\PZ^*\to4f} ={}&  
  3\Gamma^{\nu_\Pe \bar{\nu}_\Pe \nu_\mu \bar{\nu}_\mu}
+ 3\Gamma^{\Pem\Pep\mu^-\mu^+}
+ 9\Gamma^{\nu_\Pe \bar{\nu}_\Pe \mu^-\mu^+}
\notag\\&
+ 3\Gamma^{\nu_\Pe \bar{\nu}_\Pe \nu_\Pe \bar{\nu}_\Pe}
+ 3\Gamma^{\Pem\Pep\Pem\Pep}
\notag\\&
+ 6\Gamma^{\nu_\Pe \bar{\nu}_\Pe \Pu\bar{\Pu}}
+ 9\Gamma^{\nu_\Pe \bar{\nu}_\Pe \Pd\bar{\Pd}}
+ 6\Gamma^{\Pu\bar{\Pu}\Pem\Pep}
+ 9\Gamma^{\Pd\bar{\Pd}\Pem\Pep}
\notag\\&
+  \Gamma^{\Pu\bar{\Pu}\Pc\bar{\Pc}}
+ 3\Gamma^{\Pd\bar{\Pd}\Ps\bar{\Ps}}
+ 6\Gamma^{\Pu\bar{\Pu}\Ps\bar{\Ps}}
+ 2\Gamma^{\Pu\bar{\Pu}\Pu\bar{\Pu}}
+ 3\Gamma^{\Pd\bar{\Pd}\Pd\bar{\Pd}},
\notag\\[1ex]
\Gamma_{\PW\PW/\PZ\PZ\mbox{\scriptsize -interference}} ={}& 
  3\Gamma^{\nu_\Pe \Pep \Pem\bar{\nu}_\Pe}
- 3\Gamma^{\nu_\Pe \bar{\nu}_\Pe \mu^-\mu^+}
- 3\Gamma^{\nu_\Pe \Pep \mu^-\bar{\nu}_\mu}
\notag\\&
+ 2\Gamma^{\Pu\bar{\Pd}\Pd\bar{\Pu}}
- 2\Gamma^{\Pu\bar{\Pu}\Ps\bar{\Ps}}
- 2\Gamma^{\Pu\bar{\Pd}\Ps\bar{\Pc}}.
\label{eq:WW_ZZ_width}
\end{align}

\subsection{Extended Higgs models}
\label{sec:extendedmodels}

The conventions for the implementation of the SESM and THDM in
\Prophecy follow \citere{Denner:2018opp}, where the Higgs potentials
are specified and the renormalized parameters are defined.  More
details on the SESM, though with slightly different conventions, can
be found in \citeres{Altenkamp:2018bcs,Denner:2017vms}, and more
details on the THDM are provided in
\citeres{Altenkamp:2017ldc,Altenkamp:2017kxk,Denner:2017vms,Denner:2016etu}.
The renormalization of mixing angles in extended Higgs sectors has
recently been studied by various authors both in the SESM
\cite{Altenkamp:2018bcs,Denner:2017vms,Kanemura:2015fra,Bojarski:2015kra}
and the THDM
\cite{Altenkamp:2017ldc,Denner:2017vms,Denner:2016etu,Kanemura:2004mg,LopezVal:2009qy,Kanemura:2014dja,Krause:2016oke}.
{\Prophecy employs} various renormalization schemes as defined in
\citeres{Altenkamp:2017ldc,Altenkamp:2017kxk,Denner:2018opp}.

The renormalization schemes are defined via the choice of
renormalization conditions. In particular, \MSbar schemes depend on
the choice of the renormalized parameters, e.g.\ whether a mixing
angle ($\alpha$, $\beta$) or a coupling ($\lambda_{12}$, $\lambda_1$,
\ldots)
is used for renormalization, and in addition on the treatment of
tadpoles.  The renormalization schemes used in \Prophecy are based on
the tadpole schemes of \citeres{Fleischer:1980ub} and
\cite{Denner:1991kt} dubbed FJTS and PRTS, respectively, in
\citere{Denner:2018opp}. More details on these tadpole schemes can be
found in \citeres{Denner:2018opp,Denner:2016etu}.
 
\begin{sloppypar}
For a comparison of predictions based on different renormalization
schemes, a conversion of renormalized input parameters is needed.
\Prophecy automatically provides this conversion and the running of
\MSbar parameters based on the values for the scale of the input
parameters (start renormalization scale) and the scale of the
parameters used in the calculation (target renormalization scale).  In
particular, this can be used to evaluate the renormalization-scale
variation with respect to parameters of the extended Higgs sectors
renormalized in the \MSbar\ scheme (which is unrelated to the scale
dependence of $\alpha_s$).  More details on the parameter conversion
can be found in
\citeres{Altenkamp:2017ldc,Altenkamp:2018bcs,Denner:2018opp}.
\end{sloppypar}

\section{The usage of \Prophecy}
\label{se:usage}

\subsection{\Prophecy\ installation}

\Prophecy\ has been tested under various different Linux distributions
and MAC OS. It should be compilable with any standard Fortran
compiler and has been successfully tested using GNU Fortran (GCC), the
Intel Fortran Compiler, and Intel's MPI for parallel execution.

\Prophecy\ depends on 
\Collier~\cite{collier,Denner:2016kdg,Denner:2002ii,Denner:2005nn,Denner:2010tr} for
the evaluation of loop integrals.
The \Collier library ({\tt libcollier.so} or {\tt libcollier.a} for dynamic or static
linking, respectively) can be compiled
following the \Collier instructions~\cite{collier}.

For installation, download \Prophecy\ from HEPForge~\cite{prophecy}
and issue\\[1ex] 
\indent {\tt tar -xzvf Prophecy4f-3.0.tar.gz}\\
\indent {\tt cd Prophecy4f-3.0}\\
\indent {\tt make COLLIERDIR=path FC=compiler} \\[1ex]
from the command line where {\tt path} has to be the path to the
\Collier library and {\tt FC} is an optional argument that allows to
use the Fortran compiler {\tt compiler} instead of the default one.
Both variables {\tt COLLIERDIR} and {\tt FC} can also be set in the
{\tt makefile}.  The {\tt make} command generates the executable {\tt
  ./Prophecy4f}. The \Prophecy\ directory contains the {\tt
  README-3.0} file as a manual and a default input file {\tt
  defaultinput}. The source code is contained in the subdirectory {\tt
  src}, and several example runs can be found in the subdirectory {\tt
  example-runs}. The subdirectories {\tt HISTOGRAMS}, {\tt
  HISTUNWEIGHTED}, and {\tt UNWEIGHTEDEVENTS} are empty and are used
at runtime to store the results as discussed in \refse{se:unweighted}. The empty
subdirectory {\tt obj} is used to store the object files after
compilation.

\subsection{\Prophecy\ execution}

\Prophecy\ is executed using\\[1ex]
\mbox{} \hspace{1cm}
{\tt ./Prophecy4f < inputfile}\\[1ex]
where the file {\tt inputfile} specifies all the input for the current
run. If no input file is provided via standard input, \Prophecy\ does
not start.  Output is written to standard output. If an output file is
specified in the input file, most output is redirected to the output
file. Note that the path to the \Collier library has to be known
(e.g.\ by including it in the {\tt LD\_LIBRARY\_PATH} variable) if the
dynamic library is linked.

The general format of the input file is given in the default input file {\tt defaultinput}
and the input files of the subdirectories containing the example runs.
While {\tt defaultinput} specifies all relevant parameters, it is
sufficient to specify those values that differ from the default.
As a general remark, do not forget the {\tt d0} after {\tt double precision}
quantities.

\subsection{\Prophecy\ input}
\label{se:input}

In the following sections, all input options for \Prophecy\ are
discussed along with the underlying physics. We follow the structure
of the default input file {\tt defaultinput} which is provided with
the \Prophecy\ distribution.

\subsubsection{Input for calculations within the SM}
\label{sec:inputSM}

In this section, we discuss all the input that is necessary for predictions
within the Standard Model:
\begin{cpcdescription}
\item[{\bf {\tt outputfile:}}] a character string that specifies the name of the output file.
\cpcitemtable{{\tt outputfile='filename'}}{%
{\tt outputfile='\ '} &:& For a blank character string (default) all output is 
written to standard output. \\  
{\tt outputfile='filename'} &:& Any other string usable as a file name redirects the output
to a file with the given filename.}%
The plot data will be written to files named {\tt plot.*} in directory
\linebreak {\tt HISTOGRAMS}. The string {\tt plot} is replaced by the
file name of the output file if provided. Unweighted events are
written to the directory \linebreak {\tt UNWEIGHTEDEVENTS} ({\tt
  *.lhe} files) in the same manner. To allow for consistency checks,
unweighted events are also binned into distributions written to the
directory {\tt HISTUNWEIGHTED}.
\item[{\bf {\tt nevents:}}] integer that selects the number of generated weighted events.
\cpcitemtable{{\tt nevents=10000000}}{%
{\tt nevents=10000000} &:& is the default value.\\  
}%
We recommend to use at least $10^7$ events for the integrated partial decay width, 
for histograms about $5\cdot 10^7$ should be used.

\item[{\bf {\tt nunwevents:}}] integer that selects the number of
  unweighted events. 
\cpcitemtable{{\tt nunwevents=0}}{%
{\tt nunwevents=0} &:& is the default value, i.e.\ no unweighted events are produced.
}%
The unweighted events are produced in the Les Houches event file
format~\cite{Alwall:2006yp} after the generation of {\tt nevents}
weighted events to find the maximal weights used for unweighting.
Unweighted events have weight $1$ or very rarely weight $-1$. For {\tt
  nunwevents}$>$0 one has to use {\tt qsoftcoll=2} (slicing, see
below) and {\tt qrecomb=0} (no recombination, see below).  For {\tt
  nunwevents}$>$0 the two parameters are set accordingly.

\item[{\bf {\tt model:}}] integer that selects the model used for the calculation.
\cpcitemtable{{\tt model=0}}{%
{\tt model=0} &:& the Standard Model (SM) (default),\\
{\tt model=1} &:& a Higgs-Singlet Extension of the Standard Model (SESM),\\
{\tt model=2} &:& the Two-Higgs-Doublet-Model (THDM),\\
{\tt model=4} &:& the SM with a fourth fermion generation (SM4).\\
}%
Additional input that specifies the models beyond the Standard Model are discussed from
\refse{sec:sm4} to \refse{sec:thdm}.

\item[{\bf {\tt contrib:}}] integer that specifies how radiative corrections are included when 
calculating the partial decay width.
\cpcitemtable{{\tt contrib=1}}{%
  {\tt contrib=1} &:& best prediction including complete corrections,
  i.e.\  NLO corrections as defined by {\tt qqcd} (see below) and some
  higher-order effects (default),\\ 
  {\tt contrib=2} &:& use the Improved Born Approximation (IBA)
  (see \citere{Bredenstein:2006rh} for details),\\
  {\tt contrib=3} &:& calculate only      leading-order result, i.e.\ no radiative corrections are included.\\
}%
Note that the option {\tt contrib=2} is only available in the SM ({\tt
  model=0}).

\item[{\bf {\tt qqcd:}}] integer that specifies whether to use only EW
  corrections, both EW and QCD corrections, or only QCD corrections.
\cpcitemtable{{\tt qqcd=0}}{%
{\tt qqcd=0} &:& only EW corrections are included,\\
{\tt qqcd=1} &:& EW and QCD corrections are included (default),\\
{\tt qqcd=2} &:& only QCD corrections are included.\\
}%
Note that for purely leptonic final states only EW corrections contribute.

\item[{\bf {\tt qsoftcoll:}}] integer that specifies whether soft and
  collinear singularities are treated with the subtraction or the
  slicing method.
\cpcitemtable{{\tt qsoftcoll=1}}{%
{\tt qsoftcoll=1} &:& the subtraction method is used (default),\\
{\tt qsoftcoll=2} &:& the slicing method is used.\\
}%
For calculations of partial decay widths, subtraction is the preferred
option, while for the production of unweighted events slicing has to
be used.

\item[{\bf {\tt channel:}}] string that specifies the final state for which the width is to be
calculated. 
\cpcitemtable{{\tt channel= e anti-e mu anti-mu}}{%
{\tt channel= e anti-e mu anti-mu} &:& is the default.\\
}%
The final state has to be specified using {\tt e} for the electron,
{\tt mu} for the muon, {\tt nue} and {\tt num} for the corresponding
neutrinos, {\tt dq, uq, sq, cq} for the down, up, strange, and charm
quark, respectively, and the corresponding antiparticles, e.g.\ {\tt
  anti-e, anti-num, anti-dq}, etc. The four final-state particles have
to be separated by spaces.  If more than one channel is specified by
including several {\tt channel} lines in the input file, \Prophecy\ 
will calculate the different channels consecutively.  

Since final-state fermions are treated in the massless limit,
integrated partial widths usually do not differ between different
generations of fermions. For example, the integrated partial decay
width for {\tt H $\to$ e anti-e e anti-e} is the same as for {\tt H
  $\to$ mu anti-mu mu anti-mu}. Symmetric final states are an
exception, here effects of identical particles are taken into account,
i.e. {\tt H $\to$ e anti-e e anti-e} is different from {\tt H $\to$ e
  anti-e mu anti-mu}.  Moreover, in distributions (or unweighted
events) fermion-mass logarithms do show up if no photon recombination
is applied, i.e.\ fermions of different generations will in general
yield different results.  Third generation fermions cannot be used as
input.  However, the partial widths including third generation
particles like bottom quarks, tau leptons or tau neutrinos do not
differ significantly in the massless approximation from those into
fermions of the first and second generation, i.e.\ use e.g.\ {\tt H
  $\to$ mu anti-mu sq anti-sq} to calculate the {\tt H $\to$ mu
  anti-mu bq anti-bq} partial width.  Top quarks in the final state
are not supported.  The old input format of Version 1.0 is also still
supported.

In addition one can choose the special cases [see \refeq{eq:subtotal_width}
and \refeq{eq:WW_ZZ_width}]:
\cpcitemtable{{\tt channel= semi-leptonic}}{%
{\tt channel= total} &:& the total width is calculated,\\
{\tt channel= leptonic} &:& the leptonic width is calculated,\\
{\tt channel= semi-leptonic} &:& the semi-leptonic width is calculated,\\
{\tt channel= hadronic} &:& the hadronic width is calculated,\\
{{\tt channel= WW}} &:& $\Gamma_{\PH\to\PW^*\PW^*\to4f}$ is calculated,\\
{{\tt channel= ZZ}} &:& $\Gamma_{\PH\to\PZ^*\PZ^*\to4f}$ is calculated,\\
{{\tt channel= interference}} &:& $\Gamma_{\PW\PW/\PZ\PZ\mbox{\scriptsize -interference}}$ is calculated.\\
}%
If one of the above options is used, \Prophecy\ calculates the
necessary partial widths consecutively. For {\tt channel= total}, also the results for {\tt channel= WW}, {\tt channel= ZZ}, and
{\tt channel= interference} are automatically calculated.       

\item[{\bf {\tt qrecomb:}}] integer that specifies whether to use a
  recombination procedure.
\cpcitemtable{{\tt qrecomb=0}}{%
  {\tt qrecomb=0} &:& photons and fermions are not recombined,\\
  {\tt qrecomb=1} &:& the photon and the fermion with the smallest
  invariant mass are recombined if their invariant mass in GeV is smaller
  than {\tt invrecomb} (see below), i.e.\ their 4-momenta are added
  and attributed to the fermion (default).\\
}%
Note that we cannot use a proper jet-algorithm for recombination since
the lab frame of the Higgs decay is not specified;
{this would require the embedding of the Higgs decay process
into a full production process.}
For inclusive
partial widths recombination does not affect the result.  Independent
of {\tt qrecomb}, we always recombine the two QCD partons with the
smallest invariant mass in events with gluon emission to form two jets
in semileptonic decays or four jets in hadronic decays.  When
producing unweighted events for leptonic final states one has to use
{\tt qrecomb=0}, in order to create the flexibility to perform the
recombination on the event files after production.
\item[{\bf {\tt invrecomb:}}] double precision number that is used in
  the recombination procedure for {\tt qrecomb=1} (see above).
\cpcitemtable{{\tt invrecomb=5d0}}{%
{\tt invrecomb=5d0} &:& is the default, i.e.\ the photon--fermion pair
with the smallest invariant mass is recombined if the invariant mass
is smaller than $5\GeV$.\\ 
}%

\item[{\bf {\tt qrecombcolle:}}] integer that specifies whether to use a specific recombination 
procedure for electrons.
\cpcitemtable{{\tt qrecombcolle=0}}{%
{\tt qrecombcolle=0} &:& photons and electrons are not recombined inside the slicing cone,\\
{\tt qrecombcolle=1} &:& photons and electrons are recombined inside the slicing cone if {\tt qsoftcoll=2} is used (see above), i.e.\ 
their 4-momenta are added and attributed to the fermion (default).\\
}%
In the slicing approximation these recombined electron--photon pairs
are strictly collinear.  This option for recombination might be useful
to avoid large fractions of negative unweighted events for electron
final states. The idea is that photons and electrons are always
recombined within a small cone in a physical analysis, i.e.\ electrons
are used as dressed leptons in contrast to bare muons which can also
be defined without recombination.  The technical slicing cone should
always be smaller than the physical cone size, so that this
recombination is not in conflict with the physical treatment, i.e.\ 
the full flexibility of the unweighted event sample is not spoiled if
dressed electrons are assumed.  The internal slicing parameters are set
so that the ratio of the energies of an unrecombined photon and the
Higgs boson fulfils $E_\gamma/E_\PH> \delta_E =4\times10^{-4}$ and
each angle between the three-momentum of an unrecombined photon and 
the three-momentum of any
fermion is larger than $\Delta\theta = 3\times10^{-2}$.
All quantities refer to the Higgs-boson rest frame.

\item[{\bf {\tt randomseed:}}] integer that specifies how random numbers 
are generated.
\cpcitemtable{{\tt randomseed = -1}}{%
{\tt randomseed = -1} &:& use internal random numbers as in \Prophecy\ version 2.0 (default),\\
{\tt randomseed $\ge$ 0} &:& use RANLUX~\cite{Luscher:1993dy} for random number generation, 
where randomseed is used as a seed for the random number generator to obtain statistically 
independent samples.\\\phantom{aaaa}\\
}%

\end{cpcdescription}

\noindent The following input parameters can be specified with
double-precision values. Here, we also state the corresponding default
values: 
\cpctable{{\bf {\tt 1/alpha0 = 137.0359997d0 :}}}{%
{\bf {\tt mh = 125d0 }}&:& Higgs-boson mass in the SM or SM4 in GeV,\\
{\bf {\tt alphas   = 0.118d0 }}&:& strong coupling constant,\\
{\bf {\tt gf       = 1.1663787d-5 }}&:& Fermi constant $G_\mathrm{F}$ in GeV$^{-2}$,\\
{\bf {\tt mz       = 91.1876d0 }}&:& on-shell Z-boson mass in GeV,\\
{\bf {\tt mw       = 80.385d0 }}&:& on-shell W-boson mass in GeV,\\
{\bf {\tt gammaz   = 2.4952d0 }}&:& on-shell Z-boson width in GeV,\\
{\bf {\tt gammaw   = 2.085d0 }}&:& on-shell W-boson width in GeV,\\
{\bf {\tt me       = 0.510998928d-3 }}&:& electron mass in GeV,\\
{\bf {\tt mmu      = 105.6583715d-3 }}&:& muon mass in GeV,\\
{\bf {\tt mtau     = 1.77682d0      }}&:& tau mass in GeV,\\
{\bf {\tt md       = 0.100d0        }}&:& d-quark mass in GeV,\\
{\bf {\tt mu       = 0.100d0        }}&:& u-quark mass in GeV,\\
{\bf {\tt ms       = 0.100d0        }}&:& s-quark mass in GeV,\\
{\bf {\tt mc       = 1.51d0          }}&:& c-quark mass in GeV,\\
{\bf {\tt mb       = 4.92d0         }}&:& b-quark mass in GeV,\\
{\bf {\tt mt       = 172.5d0        }}&:& t-quark mass in GeV.\\
}%

The values of the fermion masses are needed, but the results are
practically independent of the specific values in the
$\alpha_{G_\mathrm{F}}$ scheme for inclusive quantities. Only if
photons and fermions are not recombined, logarithms of the fermion
masses may appear in distributions.

The gauge-boson resonances are described in the complex-mass scheme as
discussed in \refse{sec:cms}. The values of the on-shell gauge-boson
masses and widths, as given in the input, are only used to calculate
the pole masses of the gauge bosons according to \refeq{eq:m_ga_pole}.
For the actual evaluation of the Higgs decay, the gauge-boson widths
are calculated from the gauge-boson pole masses and the remaining
input as discussed in \refse{sec:cms}: If only LO results are
requested (i.e.  for {\tt contrib=3}) the LO {gauge-boson} widths are
used.  For NLO results and the IBA ({\tt contrib=1} or {\tt
  contrib=2}) we apply the NLO {gauge-boson} widths (also for the LO
sub-contribution). Note that in
\citeres{Bredenstein:2006rh,Bredenstein:2006nk,Bredenstein:2006ha} we
have presented the LO results with NLO {gauge-boson} widths. At LO,
the difference is, of course, only a higher-order effect.

In the SESM and the THDM, the NLO W and Z widths are almost identical
to the corresponding SM NLO widths.  Hence, the SM widths are used
employing a SM Higgs mass set to the mass of the decaying Higgs boson.

\subsubsection{Additional input for calculations within SM4}
\label{sec:sm4}

In \Prophecy\ a 4th fermion generation of massive fermions can be
optionally included upon setting {\tt model=4}. In the following, we
specify the additional input needed for a calculation in the model SM4. The
following options have no effect unless {\tt model=4} is used.
\begin{cpcdescription}
\item[{\bf {\tt qsm4:}}] integer that specifies which corrections are included.
\cpcitemtable{{\tt qsm4=1}}{%
  {\tt qsm4=1} &:& the full mass dependence of the additional closed
  fermion loops is taken into account at NLO, comprising the
  HWW/HZZ/HZA/HAA vertex corrections as well as all gauge-boson self-energies (default), \\
  {\tt qsm4=2} &:& in addition to the corrections used for {\tt
    qsm4=1}, the leading corrections $\propto G_\mathrm{F}^2
  m_{f,4}^4$ and $\propto \alpha_s G_\mathrm{F} m_{f,4}^2$ to the HVV
  vertices are taken in to account, which are taken from
  \citeres{Djouadi:1997rj} and \cite{Kniehl:1995ra}, respectively.
  Here, $m_{f,4}$ refers to the masses of the fourth generation
  fermions.}
\end{cpcdescription}

\noindent The masses of the two additional leptons and two additional quarks are
specified as double precision numbers, stated in the following along with their 
default values:
\cpctable{{\bf {\tt ml4      = 600d0 :}}}{%
{\bf {\tt ml4      = 600d0 }}&:& mass of the charged lepton in the 4th generation,\\
{\bf {\tt mn4      = 600d0 }}&:& mass of the neutrino in the 4th generation,\\
{\bf {\tt md4      = 600d0 }}&:& mass of the down-type quark in the 4th generation,\\
{\bf {\tt mu4      = 600d0 }}&:& mass of the up-type quark in the 4th generation.\\
}%

\subsubsection{Additional input for calculations within the SESM or THDM}
\label{sec:sesmthdm}

\Prophecy can perform calculations for the partial widths of the
CP-even neutral Higgs bosons in the SESM ({\tt model=1}) or the THDM
({\tt model=2}). Here and in the following sections, we discuss the
additional input for these models which is irrelevant for calculations
within the SM or SM4.

First of all, one has to specify the Higgs 
boson in the initial state for which the calculation has to be performed:
\begin{cpcdescription}
\item[{\bf {\tt hboson:}}] string that specifies the decaying Higgs boson.
\cpcitemtable{{\tt hboson=h0}}{%
{\tt hboson=h0} &:& light Higgs boson in the SESM or THDM (default), \\  
{\tt hboson=hh} &:& heavy Higgs boson in the SESM or THDM.}
\end{cpcdescription}

The SESM/THDM input parameters and the available renormalization
schemes are discussed in the following sections.  The SESM/THDM input
parameters of \MSbar type are defined at the renormalization scale
{\tt mrenbsm1} in the renormalization scheme specified by {\tt
  renscheme} and evolved to the renormalization scale {\tt mrenbsm2}
by solving the renormalization group equations numerically.  The
corresponding parameters at the target scale {\tt mrenbsm2} are used in the
calculation within the scheme {\tt renscheme}. In particular, this can
be used to evaluate the scale variation with respect to the BSM
\MSbar\ parameters ($\alpha_s$ is not varied).  

\subsubsection{Additional input for calculations within the SESM}
\label{sec:sesm}

\begin{sloppypar}
In this section, we discuss the input specific for the SESM (see \refse{sec:extendedmodels}). 
We follow the notation and the conventions of \citere{Denner:2018opp}.
\end{sloppypar}

\begin{cpcdescription}
\item[{\bf {\tt renscheme:}}] integer that specifies the renormalization scheme for the input parameters.
\cpcitemtable{{\tt renscheme = 5}}{%
{\tt renscheme = 0} &:& $\alpha$ \MSbar\ (running $\lambda_{12}$), i.e.\ \MSbar\ of \citere{Altenkamp:2018bcs}, \\  
{\tt renscheme = 1} &:& $\alpha$ \MSbar\ \`a la FJ (running $\lambda_{12}$), i.e.\ FJ of \citere{Altenkamp:2018bcs}, \\  
{\tt renscheme = 2} &:&  $\alpha$ on-shell (running $\lambda_{12}$),
i.e.\ OS of \citere{Denner:2018opp} for $\alpha$,\\    
{\tt renscheme = 3} &:& $\alpha$ \MSbar\ (running $\lambda_{1}$), i.e.\ \MSbar(PRTS) of \citere{Denner:2018opp}, \\  
{\tt renscheme = 4} &:& $\alpha$ \MSbar\ \`a la FJ (running $\lambda_{1}$), i.e.\ \MSbar(FJTS) of \citere{Denner:2018opp}, \\  
{\tt renscheme = 5} &:&  $\alpha$ on-shell (running $\lambda_{1}$), i.e.\ OS of \citere{Denner:2018opp} (default), \\  
{\tt renscheme = 6} &:& BFM-inspired scheme  based on Eqs.~(3.41) and (3.64) of \citere{Denner:2018opp},\\  
{\tt renscheme = 7} &:& BFM-inspired scheme  BFMS of
\citere{Denner:2018opp} based on Eqs.~(3.41) and (3.68) of \citere{Denner:2018opp}. \\  
}
\end{cpcdescription}

\noindent The masses of the Higgs bosons, the Higgs-boson mixing angle
$\alpha$, the additional Higgs-sector coupling $\lambda_{12}$, and the
renormalization scales have to be specified as double precision
numbers, stated in the following along with their default values:
\cpctable{{\bf {\tt mrenbsm1 = 125.1d0 :}}}{%
{\bf {\tt mrenbsm1 = 125.1d0 }}&:& start renormalization scale for \MSbar\ parameters (we usually use 
{\tt mrenbsm1 = mh0}),\\
{\bf {\tt mrenbsm2 = 125.1d0 }}&:& target renormalization scale for \MSbar\ parameters (we usually use
{\tt mrenbsm2 = mh0}),\\
{\bf {\tt sa       = 0.29d0 }}&:& $\sin\alpha$,\\
{\bf {\tt !ta       = 0.303d0 }}&:& {$\tan\alpha$, 
as alternative to define $\alpha$}\\
{\bf {\tt mh0      = 125.1d0 }}&:& mass of the light Higgs boson,\\
{\bf {\tt mhh      = 200d0 }}&:& mass of the heavy Higgs boson,\\
{\bf {\tt l12      = 0.07d0 }}&:& coupling $\lambda_{12}$.\\
}%
The mixing angle $\alpha$ can vary in the range
$-\pi/2 < \alpha < \pi/2$.  The line for setting {\tt ta} is an
alternative for defining the angle $\alpha$; the parameter $\alpha$,
however, should only be set once.  Our model parametrization requires
the relation $s_\alpha\lambda_{12}\ge0$ for consistency.

\subsubsection{Additional input for calculations within the THDM}
\label{sec:thdm}

\begin{sloppypar}
In this section, we discuss the input specific for the THDM (see \refse{sec:extendedmodels}). 
We follow the notation and the conventions of \citere{Denner:2018opp}.
\end{sloppypar}

\begin{cpcdescription}
\item[{\bf {\tt modeltype:}}] integer that specifies the variant of the THDM, 
as defined in \citere{Altenkamp:2017kxk}.
\cpcitemtable{{\tt modeltype = 1}}{%
{\tt modeltype = 1} &:& type I: all fermions couple to Higgs doublet
$\Phi_2$ only (default), \\
{\tt modeltype = 2} &:& type II: down-type fermions couple to $\Phi_1$, 
                         up-type fermions to $\Phi_2$, \\
{\tt modeltype = 3} &:& lepton specific: quarks couple to $\Phi_2$,
leptons to $\Phi_1$, \\
{\tt modeltype = 4} &:& flipped: down-type quarks couple to $\Phi_1$, 
                  up-type quarks and charged leptons couple to $\Phi_2$.\\
}
\item[{\bf {\tt renscheme:}}] integer that specifies the renormalization scheme for the input parameters.
\cpcitemtable{{\tt renscheme = 5}}{%
{\tt renscheme = 0} &:& $\alpha$/$\beta$ \MSbar, i.e.\ \MSbar($\alpha$) of \citeres{Altenkamp:2017ldc,Altenkamp:2017kxk}
= \MSbar(PRTS) of \citere{Denner:2018opp}, \\  
{\tt renscheme = 1} &:& $\alpha$/$\beta$ \MSbar\ \`a la FJ, i.e.\ FJ($\alpha$) of \citeres{Altenkamp:2017ldc,Altenkamp:2017kxk}
= \MSbar(FJTS) of \citere{Denner:2018opp}, \\  
{\tt renscheme = 2} &:& $\lambda_3$/$\beta$ \MSbar, i.e.\ \MSbar($\lambda_3$) of \citeres{Altenkamp:2017ldc,Altenkamp:2017kxk}, \\        
{\tt renscheme = 3} &:& $\lambda_3$/$\beta$ \MSbar \`a la FJ, i.e.\ FJ($\lambda_3$) of \citeres{Altenkamp:2017ldc,Altenkamp:2017kxk}, \\  
{\tt renscheme = 4} &:& $\alpha$/$\beta$ on-shell ($\nu_2$), i.e.\ OS2 of \citere{Denner:2018opp}, \\  
{\tt renscheme = 5} &:& $\alpha$/$\beta$ on-shell ($\nu_1$, $\nu_2$), i.e.\ OS12 of \citere{Denner:2018opp} (default), \\   
{\tt renscheme = 6} &:& $\alpha$/$\beta$ on-shell ($\nu_1$), i.e.\ OS1 of \citere{Denner:2018opp}, \\ 
{\tt renscheme = 7} &:& BFM-inspired scheme based on Eqs.~(3.41) and (3.74) of \citere{Denner:2018opp},\\  
{\tt renscheme = 8} &:& BFM-inspired scheme BFMS of
\citere{Denner:2018opp} based on Eqs.~(3.41) and (3.76) of \citere{Denner:2018opp}.  \\  
}
\end{cpcdescription}

\noindent The masses of the Higgs bosons, the mixing angles
$\alpha,\beta$, the additional Higgs-sector coupling $\lambda_5$, and the 
renormalization scales have to be specified as double precision
numbers, stated in the following along with their default values:
\cpctable{{\bf {\tt mrenbsm1 = 361d0 :}}}{%
{\bf {\tt mrenbsm1 = 361d0 }}&:& start renormalization scale for \MSbar\ parameters (we usually use \hfill\strut\hfill
{\tt mrenbsm1=(mh0+mhh+ma0+2mhp)/5}),\\
{\bf {\tt mrenbsm2 = 361d0 }}&:& target renormalization scale for
\MSbar\ parameters (we usually use \hfill\strut\hfill
{\tt mrenbsm2=(mh0+mhh+ma0+2mhp)/5}),\\
{\bf {\tt sa       = -0.355d0 }}&:& $\sin\alpha$,\\
{\bf {\tt !ta      = -0.380d0}}&:& {$\tan\alpha$, 
as alternative to define $\alpha$}\\
{\bf {\tt !cba     = 0.1d0}}&:& {$\cos(\beta-\alpha)$, 
as alternative to define $\alpha$}\\
{\bf {\tt !sgnsba  = +1}}&:& {$\mathrm{sgn}[\sin(\beta-\alpha)]$, required if {\tt cba} is input}\\
{\bf {\tt tb       = 2d0 }}&:& $\tan\beta$,\\
{\bf {\tt !sb      = 0.894d0}}&:& $\sin(\beta)$, as alternative to define $\beta$,\\ 
{\bf {\tt !cb      = 0.447d0}}&:& $\cos(\beta)$, as alternative to define $\beta$,\\ 
{\bf {\tt mh0      = 125d0 }}&:& mass of the light CP-even Higgs boson,\\
{\bf {\tt mhh      = 300d0 }}&:& mass of the heavy CP-even Higgs boson,\\
{\bf {\tt ma0      = 460d0 }}&:& mass of the CP-odd Higgs boson,\\
{\bf {\tt mhp      = 460d0 }}&:& mass of the charged Higgs boson,\\
{\bf {\tt lam5     = -1.9d0 }}&:& coupling $\lambda_{5}$.\\
}%
The mixing angles $\alpha$ and $\beta$ can vary in the ranges
$-\pi/2 < \alpha < \pi/2$ and  $0 < \beta < \pi/2$.
The lines for setting {\tt sb} and {\tt cb} 
are alternatives for defining the angle $\beta$, the ones for setting 
{\tt ta} and {\tt cba} are alternatives for defining the angle $\alpha$. 
Again the parameters $\alpha$ and $\beta$ should only be set once.
If {\tt cba} is chosen as input, the parameter {\tt sgnsba} has to be
set as well, in order to define $\alpha$ uniquely.

\subsection{Parallel execution using the MPI standard}
\label{se:mpi}

\Prophecy\ supports parallel execution using MPI. To use the parallel
version, one has to compile the program using the preprocessor flag
{\tt -Dmpiuse} and make sure that proper MPI libraries are linked.
The program will produce {\tt nevents} weighted events in total and
{\tt nunwevents} unweighted events per core.  The parallel version of
\Prophecy\ has been tested using Intel's Fortran compiler with Intel's
MPI.

\section{\Prophecy\ output and sample runs}
\label{se:output}

All output by \Prophecy\ is written to standard output or the output
file specified in the input. Information about the \Collier library is
always written to standard output and the \Collier output directory.
In the output, the model under consideration is given along with the
initial-state Higgs boson and the four-fermion final state for which
the partial width and corresponding differential distributions are
calculated. The input parameters are listed along with derived
parameters, in particular the gauge-boson pole masses and widths (see
\refse{sec:inputSM}).

For the SESM and the THDM, also information to define the model input
is given, i.e.\ the renormalization scheme and the renormalization
scale (see \refses{sec:sesm} and \ref{sec:thdm}).  For convenience,
\Prophecy\ also converts the employed model parameters to the
corresponding values of the model parameters in other
renormalization schemes, based on two different parameter conversion
techniques (see
\citeres{Altenkamp:2017ldc,Altenkamp:2017kxk,Altenkamp:2018bcs,Denner:2018opp}),
and provides them in the output.

The relevant options, which are available as input (see
\refse{sec:inputSM}), are also listed, in particular the number of
requested events, the radiative corrections included in the
calculation and the options for lepton--photon recombination.

During the Monte Carlo integration, \Prophecy\ provides intermediate
results for the integrated partial width under consideration which allows
one to monitor the progress of the calculation.

Once the calculation is finished, the full result containing all
radiative corrections is provided in the output along with an estimate
of the Monte Carlo integration error. Subcontributions of the full
result such as the LO result or the EW and QCD corrections are also
given if they are included and non-zero.

If unweighted events are requested, the output includes information
how many unweighted events have been already produced. After the
production is finished, we include information if and how often the
largest weight used for the unweighting procedure has been exceeded.
If this number becomes large, the number of weighted events {\tt
  nevents} should be increased. We also give the number of unweighted
events with a negative weight. If this number exceeds a few percent
for final states with electrons, one should consider to use {\tt
  qrecombcolle=1} (see \refse{sec:inputSM}).

If more than one final state is requested in the input file,
\Prophecy\ lists all requested channels at the beginning of the
output. Then, the output for the different final states is provided
consecutively as the calculation is performed in complete analogy to a
run with only one final state. After finishing the calculation for all
requested final states, a summary of all results is printed.

If {\tt channel=total}, {\tt channel=leptonic}, {\tt
  channel=semi-leptonic}, or \linebreak {\tt channel=hadronic} is used
to request the corresponding width of the Higgs boson, all required
four-fermion final states are listed in the beginning.  Output for
each channel is created along with the calculation. In the end, the
corresponding width is calculated according to Eqs.~\refeq{eq:total_width}
and \refeq{eq:subtotal_width}. For {\tt channel=total}, i.e.\ the
Higgs-boson width for decays into all four-fermion final states, we
additionally give the Higgs-boson width for the decays $\PH\to\PW\PW$
and $\PH\to\PZ\PZ$ along with the corresponding interference
contribution as defined in Eq.~\refeq{eq:WW_ZZ_width}.

\subsection{Unweighted events}
\label{se:unweighted}

Unweighted events are written to the directory {\tt UNWEIGHTEDEVENTS}
in the Les Houches event file format ({\tt
  *.lhe})~\cite{Alwall:2006yp}. For each run, we provide Born-level
unweighted events and unweighted events including radiative
corrections.  These files contain also the complete output in their
headers.  As a cross check the unweighted events are binned into
distributions (see \refse{se:Differential}) and written to the directory
{\tt HISTUNWEIGHTED}. These histograms are equivalent to histograms
obtained by binning the events in the {\tt *.lhe} files accordingly.

Note that the unweighted events are not suitable for a subsequent parton-shower simulation for 
multi-photon emissions from the Higgs-boson decay products. Showering the unweighted events
provided by \Prophecy\ would lead to a partial double counting of the photon-emission from the 
final-state leptons.

In unweighting runs with more than $10^6$ unweighted events in a single run it is
possible that fewer unweighted events are generated than requested. This is 
caused by a 32bit integer overflow and can be solved by using 64bit integers 
everywhere. The default size of all integers can be controlled using the 
following compiler options for the Fortran compiler:\\[1ex]
\hspace*{1cm} \begin{tabular}{rl}
  gfortran: & {\tt -fdefault-integer-8}\\
  ifort: & {\tt -integer-size 64}\\
\end{tabular}\\[1ex]
These options are the default in the supplied makefile.

\subsection{Differential distributions}
\label{se:Differential}

For leptonic or semi-leptonic final states, a few default histograms
corresponding to the distributions presented in
\citeres{Bredenstein:2006rh,Bredenstein:2006nk,Bredenstein:2006ha} are
produced in the directory {\tt HISTOGRAMS}. They can be modified in
the subroutine {\tt create\_histo} in the file {\tt src/public.F}.
There, a subroutine called {\tt histogram} is called. Its first two
parameters correspond to the range of the histogram, the third
parameter to the variable of the distribution, and the number 50
refers to the number of bins.  The output format of the histograms is
detailed in the corresponding output files.

If one is interested in differential distributions it is not useful to
specify more than one decay channel for a given \Prophecy\ run. The
results for one channel would be simply overwritten by the next
channel.

\begin{sloppypar}
By default, \Prophecy\ provides histograms for invariant masses around
the W- and Z-boson resonances. The invariant masses refer to the first
two or the last two particles listed in the output for the final state
under consideration, as indicated by the file names.  Note that the
particle ordering can vary between the input file and the output since
particles are always ordered so that the first two and the last two
particles correspond to the gauge-boson resonances. Note that the
distributions do not necessarily have direct physical significance if
identical or invisible particles are present in the final state. The
following invariant-mass distributions are provided by default:
\cpctable{{\tt outputfile.inv12.6010$\,\,$}}{
  {\tt outputfile.inv12.5090}     &&    inv.~mass $m_{12}$ between 50 and $90 \GeV$,\\
  {\tt outputfile.inv12.7585}     &&    inv.~mass $m_{12}$ between 75 and $85 \GeV$,\\
  {\tt outputfile.inv12.60100}    &&    inv.~mass $m_{12}$ between 60 and $100 \GeV$,\\
  {\tt outputfile.inv12.8595}     &&    inv.~mass $m_{12}$ between 85 and $95 \GeV$,\\
  {\tt outputfile.inv34.5090}     &&    inv.~mass $m_{34}$ between 50 and $90 \GeV$,\\
  {\tt outputfile.inv34.7585}     &&    inv.~mass $m_{34}$ between 75 and $85 \GeV$,\\
  {\tt outputfile.inv34.60100}    &&    inv.~mass $m_{34}$ between 60 and $100 \GeV$,\\
  {\tt outputfile.inv34.8595} && inv.~mass $m_{34}$ between 85 and $95\GeV$,  } 
where $m_{ij}$ denotes the invariant mass of the fermion
pair $f_i\bar{f}_j$.
\end{sloppypar}

In addition, the following distributions are available,
where again the particle numbering refers the to the final state as
printed in the output and, for example, $k_3$ denotes the four-momentum
of the third particle:

\cpctable{{\tt outputfile.phitrf2f3}}{ {\tt outputfile.cthv2f2} && the
  cosine of the angle between $(k_3+k_4)$ and $k_2$ in the Higgs rest
  frame, see e.g. Fig.~12 in \citere{Bredenstein:2006ha} (however, the
  particle
  numbering is different there),\\
  {\tt outputfile.cthv2f3} && the cosine of the angle of $k_3$ with
  respect to $(k_3+k_4)$ in the $(k_3+k_4)$
  rest frame, see e.g. Fig.~14 in \citere{Bredenstein:2006rh},\\
  {\tt outputfile.phitrf2f3} && angle between particle 2 and 3 in the
  transverse plane according to Fig.~15 in
  \citere{Bredenstein:2006rh},\\
  {\tt outputfile.cthf1f3} && the cosine of the angle between particle
  1 and 3 according to Fig.~16
  in \citere{Bredenstein:2006rh},\\
  {\tt outputfile.phi} && the angle between the decay planes according
  to Eq.~(7.9) of \citere{Bredenstein:2006rh} (only provided for fully leptonic
  final states),\\
  {\tt outputfile.cphihad} && the absolute value of the cosine of the
  angle between the decay planes (one plane spanned by particles 1 and
  2, the second plane spanned by particles 3 and 4), as defined in Eq.
  (4.2) of
  \citere{Bredenstein:2006ha} (only provided for semi-leptonic final states).\\
  }

The present version of \Prophecy\ does not provide histograms for
hadronic final states.

\subsection{Sample runs}

Examples for input files and the resulting output are given in the
directory {\tt example-runs}. It contains subdirectories 
which are discussed in the following:

{\tt example-paper}:\\*
This directory contains input files for the decay modes $\PH\to\PZ\PZ
\to \Pem\Pep\mu^-\mu^+$, $\PH\to\PZ\PZ\to\Pem\Pep\Pem\Pep$,
$\PH\to\PW\PW\to\nu_\Pe\Pep\mu^-\bar\nu_\mu$, and
$\PH\to\PW\PW\to\nu_\Pe\Pep\Pem\bar\nu_\Pe$ 
{in the SM} with Higgs masses of
$140$, $170$, and $200\GeV$ for the input parameters of
\citere{Bredenstein:2006rh}.  For reference the corresponding output
files are provided in {\tt out.*} and the histograms in the directory
{\tt example-paper/HISTOGRAMS}. The results for the WW-mediated
channels differ by up to 0.5\% from those given in Table~1 of
\citere{Bredenstein:2006rh} due to a bug in the renormalization of the
(complex) W-boson mass which has been removed in the meanwhile. The
ZZ-mediated channels give slightly different results from those in
\citeres{Bredenstein:2006rh,Bredenstein:2006nk,Bredenstein:2006ha}
since the top-mass effects in the Z width calculation are treated in
an improved manner in the recent version of \Prophecy.

{\tt example-channels}:\\*
This directory contains input and corresponding output files for
all final states in the SM for a Higgs mass of $125\GeV$ for the default input
parameter set.

{\tt example-unweighted}:\\*
This directory contains input and corresponding output files for the
production of unweighted events for leptonic final states in the SM, a
Higgs mass of $125\GeV$, and the default input parameter set along with
the corresponding distributions.  The files with the unweighted events
are not part of the distribution due to their size.

{\tt example-SESM}:\\*
This directory contains input files for the decay mode $\Ph\to\PW\PW
\to  \nu_\mu \mu^+ \Pem \bar{\nu}_\Pe$  of the light CP-even Higgs boson of the
SESM scenario BHM200 of \citeres{Altenkamp:2018bcs,Denner:2018opp}.

{\tt example-THDM}:\\*
This directory contains input files for the decay mode
$\Ph\to\PW\PW\to \nu_\mu \mu^+  \Pem \bar{\nu}_\Pe $ of the light CP-even Higgs
boson of the THDM scenario Aa of
\citeres{Altenkamp:2017ldc,Altenkamp:2017kxk}, which is identical to
A1 of \citere{Denner:2018opp}.

\section{Conclusions}
\label{se:conclusion}

The Monte Carlo program \Prophecy\ calculates predictions for
Higgs-boson decays into four-fermion final states via
$\PH\to\PW\PW/\PZ\PZ\to4\Pf$ including the full set of next-to-leading
order corrections of the strong and electroweak interactions. In
addition to predictions within the Standard Model, \Prophecy\ provides
{the partial decay widths of the CP-even, neutral Higgs bosons}
in several extensions of the Standard
Model, i.e.\ the Standard Model extended by a fourth fermion
generation, a simple Higgs-singlet extension of the SM, and the
Two-Higgs-Doublet Model. 
{For these SM extensions, \Prophecy supports different types of
renormalization schemes based on $\MSbar$, on-shell, or
symmetry-inspired renormalization conditions and different variants thereof.
This allows for important checks on the perturbative stability and, thus, reliability
of predictions by estimating residual
renormalization scale and renormalization scheme dependences.}

{In the past, state-of-the-art predictions for SM Higgs decay widths
have been produced by the program {\sc HDECAY} in tandem with \Prophecy.
With the new versions of the two programs, both supporting various common
renormalization schemes, uniform predictions within the THDM and SESM
become possible, which is an important step towards Higgs precision
physics in models with extended Higgs sectors.}

\section*{Acknowledgements}
We are indebted to A.~Bredenstein and M.M.~Weber for constructing the
first versions of \Prophecy. 
We thank L.~Altenkamp, M. Boggia, J.N.~Lang, and H. Rzehak for their
contributions to the extension of \linebreak \Prophecy for extended Higgs sectors.
A.D.\ acknowledges financial support by the
German Research Foundation (DFG) under reference number DE 623/5-1.
{S.D.\ gratefully acknowledges support by the 
German Bundesministerium f\"ur Bildung und Forschung (BMBF) under contract
no.~05H18VFCA1.} 
A.M. is supported in part by the DFG through the CRC/Transregio 
"P3H: Particle Physics Phenomenology after the Higgs Discovery" (TRR257).






\begin{thebibliography}{0}
\bibitem{xBredenstein:2006rh}
  A.~Bredenstein, A.~Denner, S.~Dittmaier and M.~M.~Weber,
  Phys.\ Rev.\ D {\bf 74} (2006) 013004
  [hep-ph/0604011].

\bibitem{xBredenstein:2006nk}
  A.~Bredenstein, A.~Denner, S.~Dittmaier and M.~M.~Weber,
  Nucl.\ Phys.\ Proc.\ Suppl.\  {\bf 160} (2006) 131
  [hep-ph/0607060].

\bibitem{xBredenstein:2006ha}
  A.~Bredenstein, A.~Denner, S.~Dittmaier and M.~M.~Weber,
  JHEP {\bf 0702} (2007) 080
  [hep-ph/0611234].

\bibitem{xAltenkamp:2017ldc}
  L.~Altenkamp, S.~Dittmaier and H.~Rzehak,
  JHEP {\bf 1709} (2017) 134
  [arXiv:1704.02645 [hep-ph]].

\bibitem{xAltenkamp:2017kxk}
  L.~Altenkamp, S.~Dittmaier and H.~Rzehak,
  JHEP {\bf 1803} (2018) 110
  [arXiv:1710.07598 [hep-ph]].

\bibitem{xAltenkamp:2018bcs}
  L.~Altenkamp, M.~Boggia and S.~Dittmaier,
  JHEP {\bf 1804} (2018) 062
  [arXiv:1801.07291 [hep-ph]].

 \bibitem{xDenner:2018opp}
  A.~Denner, S.~Dittmaier and J.~N.~Lang,
  JHEP {\bf 1811} (2018) 104
  [arXiv:1808.03466 [hep-ph]].
                               
\end{thebibliography}

\begin{thebibliography}{99}
\frenchspacing
\newcommand{\epj}[3]{{\sl Eur. Phys. J.} {\bf #1} (19#2) #3}
\newcommand{\zp}[3]{{\sl Z. Phys.} {\bf #1} (19#2) #3}
\newcommand{\np}[3]{{\sl Nucl. Phys.} {\bf #1} (19#2) #3}
\newcommand{\phm}[3]{{\sl Phil. Mag.} {\bf #1} (19#2) #3}
\newcommand{\pl}[3]{{\sl Phys. Lett.} {\bf #1} (19#2) #3}
\newcommand{\pr}[3]{{\sl Phys. Rev.} {\bf #1} (19#2) #3}
\newcommand{\prep}[3]{{\sl Phys.\ Rep.} {\bf #1} (19#2) #3}
\newcommand{\prl}[3]{{\sl Phys. Rev. Lett.} {\bf #1} (19#2) #3}
\newcommand{\prs}[3]{{\sl Proc. Roy. Soc.} {\bf #1} (19#2) #3}
\newcommand{\fp}[3]{{\sl Fortschr. Phys.} {\bf #1} (19#2) #3}
\newcommand{\cpc}[3]{{\sl Comput. Phys. Commun.} {\bf #1} (19#2) #3}
\newcommand{\ijmp}[3]{{\sl Int. J. Mod. Phys.} {\bf #1} (19#2) #3}
\newcommand{\nim}[3]{{\sl Nucl. Instr. Meth.} {\bf #1} (19#2) #3}
\newcommand{\nc}[3]{{\sl Nuovo Cimento} {\bf #1} (19#2) #3}
\newcommand{\vj}[4]{{\sl #1} {\bf #2} (19#3) #4}
\newcommand{\jcp}[3]{{\sl J. Comp. Phys.} {\bf #1} (19#2) #3}

\bibitem{Bredenstein:2006rh}
  A.~Bredenstein, A.~Denner, S.~Dittmaier and M.~M.~Weber,
  Phys.\ Rev.\ D {\bf 74} (2006) 013004
  [hep-ph/0604011].

\bibitem{Bredenstein:2006nk}
  A.~Bredenstein, A.~Denner, S.~Dittmaier and M.~M.~Weber,
  Nucl.\ Phys.\ Proc.\ Suppl.\  {\bf 160} (2006) 131
  [hep-ph/0607060].

\bibitem{Bredenstein:2006ha}
  A.~Bredenstein, A.~Denner, S.~Dittmaier and M.~M.~Weber,
  JHEP {\bf 0702} (2007) 080
  [hep-ph/0611234].

\bibitem{Altenkamp:2017ldc}
  L.~Altenkamp, S.~Dittmaier and H.~Rzehak,
  JHEP {\bf 1709} (2017) 134
  [arXiv:1704.02645 [hep-ph]].

\bibitem{Altenkamp:2017kxk}
  L.~Altenkamp, S.~Dittmaier and H.~Rzehak,
  JHEP {\bf 1803} (2018) 110
  [arXiv:1710.07598 [hep-ph]].

\bibitem{Altenkamp:2018bcs}
  L.~Altenkamp, M.~Boggia and S.~Dittmaier,
  JHEP {\bf 1804} (2018) 062
  [arXiv:1801.07291 [hep-ph]].

\bibitem{Denner:2018opp}
  A.~Denner, S.~Dittmaier and J.~N.~Lang,
  JHEP {\bf 1811} (2018) 104
  [arXiv:1808.03466 [hep-ph]].

\bibitem{Dittmaier:2011ti}
  S.~Dittmaier, C.~Mariotti, G.~Passarino and R.~Tanaka {\it et al.} [LHC Higgs Cross Section Working Group],
  CERN-2011-002, arXiv:1101.0593 [hep-ph].

\bibitem{Dittmaier:2012vm}
  S.~Dittmaier, C.~Mariotti, G.~Passarino and R.~Tanaka {\it et al.} [LHC Higgs Cross Section Working Group],
  CERN-2012-002, arXiv:1201.3084 [hep-ph].

\bibitem{Heinemeyer:2013tqa}
  S.~Heinemeyer, C.~Mariotti, G.~Passarino and R.~Tanaka {\it et al.}  [LHC Higgs Cross Section Working Group],
  CERN-2013-004, arXiv:1307.1347 [hep-ph].

\bibitem{deFlorian:2016spz}
  D.~de Florian {\it et al.} [LHC Higgs Cross Section Working Group],
  CERN-2017-002-M, arXiv:1610.07922 [hep-ph].

\bibitem{Djouadi:1997yw}
  A.~Djouadi, J.~Kalinowski and M.~Spira,
  Comput.\ Phys.\ Commun.\  {\bf 108} (1998) 56
  [hep-ph/9704448].


\bibitem{Djouadi:2006bz}
  A.~Djouadi, M.~M.~M{\"u}hlleitner and M.~Spira,
  Acta Phys.\ Polon.\ B {\bf 38} (2007) 635
  [hep-ph/0609292].

\bibitem{Djouadi:2018xqq}
  A.~Djouadi, J.~Kalinowski, M.~M{\"u}hlleitner and M.~Spira,
  Comput.\ Phys.\ Commun.\  {\bf 238} (2019) 214
  [arXiv:1801.09506 [hep-ph]].

\bibitem{Krause:2018wmo}
  M.~Krause, M.~M{\"u}hlleitner and M.~Spira,
  Comput.\ Phys.\ Commun.\  {\bf 246} (2020) 106852
  [arXiv:1810.00768 [hep-ph]].

\bibitem{Contino:2014aaa}
  R.~Contino, M.~Ghezzi, C.~Grojean, M.~M{\"u}hlleitner and M.~Spira,
  Comput.\ Phys.\ Commun.\  {\bf 185} (2014) 3412
  [arXiv:1403.3381 [hep-ph]].

\bibitem{Gunion:2002zf}
  J.~F.~Gunion and H.~E.~Haber,
  Phys.\ Rev.\ D {\bf 67} (2003) 075019
  [hep-ph/0207010].

\bibitem{Branco:2011iw}
  G.~C.~Branco, P.~M.~Ferreira, L.~Lavoura, M.~N.~Rebelo, M.~Sher and J.~P.~Silva,
  Phys.\ Rept.\  {\bf 516} (2012) 1
  [arXiv:1106.0034 [hep-ph]].

\bibitem{Schabinger:2005ei}
  R.~M.~Schabinger and J.~D.~Wells,
  Phys.\ Rev.\ D {\bf 72} (2005) 093007
  [hep-ph/0509209].

\bibitem{Patt:2006fw}
  B.~Patt and F.~Wilczek,
  hep-ph/0605188.

\bibitem{Bowen:2007ia}
  M.~Bowen, Y.~Cui and J.~D.~Wells,
  JHEP {\bf 0703} (2007) 036
  [hep-ph/0701035].

\bibitem{collier} 
A.~Denner, S.~Dittmaier, and L.~Hofer, Collier,\\
\href{https://collier.hepforge.org/}{\url{https://collier.hepforge.org/}}.

\bibitem{Denner:2016kdg}
  A.~Denner, S.~Dittmaier and L.~Hofer,
  Comput.\ Phys.\ Commun.\  {\bf 212} (2017) 220
  [arXiv:1604.06792 [hep-ph]].

\bibitem{Denner:2011vt}
  A.~Denner, S.~Dittmaier, A.~M{\"u}ck, G.~Passarino, M.~Spira, C.~Sturm, S.~Uccirati and M.~M.~Weber,
  Eur.\ Phys.\ J.\ C {\bf 72} (2012) 1992
  [arXiv:1111.6395 [hep-ph]].


\bibitem{Boselli:2015aha}
  S.~Boselli, C.~M.~Carloni Calame, G.~Montagna, O.~Nicrosini and F.~Piccinini,
  JHEP {\bf 1506} (2015) 023
  [arXiv:1503.07394 [hep-ph]].

\bibitem{Boselli:2017pef}
  S.~Boselli, C.~M.~Carloni Calame, G.~Montagna, O.~Nicrosini, F.~Piccinini and A.~Shivaji,
  JHEP {\bf 1801} (2018) 096
  [arXiv:1703.06667 [hep-ph]].

\bibitem{Denner:1999gp}
  A.~Denner, S.~Dittmaier, M.~Roth and D.~Wackeroth,
  Nucl.\ Phys.\ B {\bf 560} (1999) 33
  [hep-ph/9904472].

\bibitem{Denner:2005fg}
  A.~Denner, S.~Dittmaier, M.~Roth and L.~H.~Wieders,
  Nucl.\ Phys.\ B {\bf 724} (2005) 247
   [Erratum-ibid.\ B {\bf 854} (2012) 504]
  [hep-ph/0505042].

\bibitem{Denner:2017vms}
  A.~Denner, J.~N.~Lang and S.~Uccirati,
  JHEP {\bf 1707} (2017) 087
  [arXiv:1705.06053 [hep-ph]].

\bibitem{Denner:2016etu}
  A.~Denner, L.~Jenniches, J.~N.~Lang and C.~Sturm,
  JHEP {\bf 1609} (2016) 115
  [arXiv:1607.07352 [hep-ph]].

\bibitem{Kanemura:2015fra}
  S.~Kanemura, M.~Kikuchi and K.~Yagyu,
  Nucl.\ Phys.\ B {\bf 907} (2016) 286
  [arXiv:1511.06211 [hep-ph]].

\bibitem{Bojarski:2015kra}
  F.~Bojarski, G.~Chalons, D.~Lopez-Val and T.~Robens,
  JHEP {\bf 1602} (2016) 147
  [arXiv:1511.08120 [hep-ph]].

\bibitem{Kanemura:2004mg}
  S.~Kanemura, Y.~Okada, E.~Senaha and C.-P.~Yuan,
  Phys.\ Rev.\ D {\bf 70} (2004) 115002
  [hep-ph/0408364].

\bibitem{LopezVal:2009qy}
  D.~Lopez-Val and J.~Sola,
  Phys.\ Rev.\ D {\bf 81} (2010) 033003
  [arXiv:0908.2898 [hep-ph]].

\bibitem{Kanemura:2014dja}
  S.~Kanemura, M.~Kikuchi and K.~Yagyu,
  Phys.\ Lett.\ B {\bf 731} (2014) 27
  [arXiv:1401.0515 [hep-ph]].

\bibitem{Krause:2016oke}
  M.~Krause, R.~Lorenz, M.~M{\"u}hlleitner, R.~Santos and H.~Ziesche,
  JHEP {\bf 1609} (2016) 143
  [arXiv:1605.04853 [hep-ph]].

\bibitem{Fleischer:1980ub}
  J.~Fleischer and F.~Jegerlehner,
  Phys.\ Rev.\ D {\bf 23} (1981) 2001.

\bibitem{Denner:1991kt}
  A.~Denner,
  Fortsch.\ Phys.\  {\bf 41} (1993) 307
  [arXiv:0709.1075 [hep-ph]].



\bibitem{Denner:2002ii}
  A.~Denner and S.~Dittmaier,
  Nucl.\ Phys.\ B {\bf 658} (2003) 175
  [hep-ph/0212259].
  
\bibitem{Denner:2005nn}
  A.~Denner and S.~Dittmaier,
  Nucl.\ Phys.\ B {\bf 734} (2006) 62
  [hep-ph/0509141].
  
\bibitem{Denner:2010tr}
  A.~Denner and S.~Dittmaier,
  Nucl.\ Phys.\ B {\bf 844} (2011) 199
  [arXiv:1005.2076 [hep-ph]].

\bibitem{prophecy} 
A.~Denner, S.~Dittmaier, and A.~M\"uck, \Prophecy,\\
\href{https://prophecy4f.hepforge.org/}{\url{https://prophecy4f.hepforge.org/}}.

\bibitem{Luscher:1993dy}
  M.~L\"uscher,              
  Comput.\ Phys.\ Commun.\  {\bf 79} (1994) 100
  [hep-lat/9309020].

\bibitem{Djouadi:1997rj}
  A.~Djouadi, P.~Gambino and B.~A.~Kniehl,
  Nucl.\ Phys.\ B {\bf 523} (1998) 17
  [hep-ph/9712330].
  
\bibitem{Kniehl:1995ra}
  B.~A.~Kniehl,
  Phys.\ Rev.\ D {\bf 53} (1996) 6477
  [hep-ph/9602304].

\bibitem{Alwall:2006yp}
  J.~Alwall {\it et al.},
  Comput.\ Phys.\ Commun.\  {\bf 176} (2007) 300
  [hep-ph/0609017].
  
\end{thebibliography}







\end{document}